# Identifying predictive biomarkers of CIMAvaxEGF success in advanced Lung Cancer Patients

Patricia Luaces[1], Lizet Sanchez[1], Danay Saavedra[1], Tania Crombet[1], Wim Van der Elst[2], Ariel Alonso[2], Geert Molenberghs[2], Agustin Lage[1]

*Abstract—* **Objectives:** To identify predictive biomarkers of CIMAvaxEGF success in the treatment of Non–Small Cell Lung Cancer Patients. **Methods:** Data from a clinical trial evaluating the effect on survival time of CIMAvax-EGF versus best supportive care were analysed retrospectively following the causal inference approach. Pre-treatment potential predictive biomarkers included basal serum EGF concentration, peripheral blood parameters and inmunocenescence biomarkers (The proportion of CD8 + CD28- T cells, CD4+ and CD8+ T cells, CD4/CD8 ratio and CD19+ B cells. The 33 patients with complete information were included. The predictive causal information (PCI) was calculated for all possible models. The model with a minimum number of predictors, but with high prediction accuracy (PCI>0.7) was selected. Good, rare and poor responder patients were identified using the predictive probability of treatment success. **Results:** The mean of PCI was increase from 0.486, when only one predictor is considered, to 0.98 using the multivariate approach with all predictors. The model considering the proportion of CD4+ T cell, basal EGF concentration, NLR, Monocytes and Neutrophils as predictors was selected (PCI>0.74). Patients predicted as good responder according to the pre-treatment biomarkers values treated with CIMAvax-EGF had a significant higher observed survival compared with control group (p=0.03). No difference was observed for bad responders. **Conclusions:** Peripheral blood parameters and inmunocenescence biomarkers together with basal EGF concentration in serum resulted good predictors of the CIMAvax-EGF success in advanced NSCLC. The study illustrates the application of a new methodology, based on causal inference, to evaluate of multivariate pre-treatment predictors.

## I. Introduction

CANCER natural history involves interactions between tumor and host defense mechanisms. The therapeutic potential of host-specific and tumor-specific immune responses is well recognized and, after many years, immunotherapies directed at inducing or augmenting these responses are entering clinical practice. In particular, the epidermal growth-factor receptor superfamily is an attractive therapeutic target because it is commonly overexpressed in malignant disease, regulates many vital cellular processes, and seems to be a negative prognostic indicator. CIMAvax-EGF is a therapeutic anticancer vaccine, developed in Cuba under the concept that inducing EGF deprivation, which involves manipulating an individual's immune response to release its own effector antibodies (Abs) against EGF, tumor size or its progression can be reduced.

CIMAvax-EGF demonstrated to be safe and immunogenic in advanced non-small cell lung cancer (NSCLC) patients in several clinical trials [1-5]. However, there is evidence of heterogeneous response to the vaccine. Patients with short-term and long-term survival were differentiated between those treated with CIMAvax-EGF [6]. In the phase II and phase III trials conducted, patient developed a ''good antibody response'' (anti-EGF antibody titers ≥ 1:4,000 sera dilution) seemed to have significantly better survival compared with patients who had lower anti-EGF antibody responses[1, 3, 4]. On the other hand, correlation between EGF concentration at baseline and length of survival was observed since the phase I study [5]. The subsequent studies corroborated also this fact, vaccinated patients with serum basal EGF concentration >870 pg/ml showed larger survival as compared with controls with the same EGF serum level [1, 2]. Furthermore, immunosenescence markers as the proportion of CD8+CD28− cells, CD4 cells, and the CD4/CD8 ratio after first-line chemotherapy was also associated with CIMAvax-EGF clinical benefit. All these studies point the importance given to the search of predictive biomarkers that allow the selection of patients who can receive a real benefit with the vaccine.

Although several attempts have been done to find predictive biomarkers of clinical benefit of CIMAvax-EGF, always each potential predictor was evaluated separately. The univariate approach used has the advantage that is easy to interpret and use simple statistical techniques, comprehensible to the medical community. Nevertheless, a multivariate approach gives a much richer and realistic picture than looking at a single variable and provides a powerful test of significance to validate biomarkers compared to univariate techniques. Multivariate approach allows researchers to look at relationships between variables in an overarching way.

The aim of this study is to evaluate multivariate predictors of CIMAvax-EGF therapeutic success using the causal inference approach.

## II. METHODS

### A. Data

We analyzed data from patients with histologic evidence of Non-Small Cell Lung Cancer (NSCLC) stage IIIb-IV recruited for a controlled phase III trial (http://www.who.int/ictrp/network/rpcec/en/; Cuban Public Registry of Clinical Trials; Trial number RPCEC00000161).

[1] Center of Molecular Immunology, Havana, AP. 16040 Cuba (phone: 053-2143145; e-mails: Patricia LORENZO LUACES: patricial@cim.sld.cu, Lizet SANCHEZ: lsanchez@cim.sld.cu ).

[2] Interuniversity Institute for Biostatistics and statistical Bioinformatics (I-BioStat); e-mails: Ariel ALONSO ABAD ariel.alonsoabad@kuleuven.be ; Geert MOLENBERGHS: geert.molenberghs@uhasselt.be

We selected all patients with measures of pre-treatment basal EGF concentration, peripheral blood parameters, inflammation and inmunocenescence biomarkers. The methods and results of these trials have been reported elsewhere[1, 7]. Briefly, patients were randomized to either vaccine Arm (CIMAvaxEGF plus Best Supportive Care) or Control Arm (only Best Supportive Care). The eligible patients were those aged 18 years or older with histologically or cytological confirmed stage IIIb or IV NSCLC, and with an Eastern Cooperative Oncology Group (ECOG) performance status of 0 to 2. All patients had received 4 to 6 cycles of platinum-based chemotherapy before random assignment and had finished first-line chemotherapy at least 4 weeks before entering in the trial. Pregnancy or lactation, secondary malignancies, or history of hypersensitivity to foreign proteins rendered patients ineligible. The primary efficacy endpoint was the survival time, defined as elapsed time since trial inclusion to death.

The potential pre-treatment predictive variables considered were basal serum EGF concentration, peripheral blood populations (absolute neutrophils, lymphocyte, monocytes and platelets counts, neutrophil-to-lymphocyte ratio and platelet-to-lymphocyte ratio) and inmunocenescence biomarkers (The proportion of CD4 + T cells and CD4/CD8 ratio). We only included in this work data from 40 patients who had completed measures of the potential pre-treatment predictive variables..

*B. Modeling approach*

*Calculation of the predictive causal inference association for all possible models*

Following the causal inference approach proposed by Alonso and colleagues [8], we analyzed each of our potential predictors separately, in a univariate way, and later all possible combinations of them. For all, the predictive causal information (PCI) was calculated. It was defined as the correlation between the treatment effect and the predictors. PCI indicate the prediction accuracy, i.e., how accurately one can predict the individual causal treatment effect on the true endpoint for a given individual, using his pre-treatment predictor measurements. The interpretation is similar to the widely used correlation coefficients. If PCI is exactly 1, that's indicate a perfect prediction of the individual causal treatment effect using the values of predictors. The closer the values are to zero, the lower the model's ability to predict the real benefit of the patient from the values of the predictors. The prediction accuracy was classified according to the value of the PCI as negligible (PCI≤0.3), low accurate (0.3< PCI≤0.5), moderately accurate (0.5< PCI≤0.7), highly accurate (0.7< PCI ≤0.9) and very high (0.9< PCI ≤1). All calculation where performed using the R library EffectTreat.

Selection of a model taking into account its complexity and prediction accuracy

The inclusion of more predictors will always lead to an increase in information about the effect of individual causal treatment. However, measuring and collecting data on multiple predictors can increase the burden for clinical investigators, patients and generate higher costs. We propose to follow the criterion of parsimony, that is, to select a model with the correct amount of predictors necessary to explain the data well. Firstly, within the combinations with the same number of predictors, we select the one with the highest PCI value. Then we classify its accuracy according to the scale previously described. Finally, we chose the model with a minimum number of predictors (lowest complexity), but with all the PCI values above 0.7, that is, with a high prediction accuracy.

*Identification of good, rare and bad responders to the treatment*

The classic definition of responder (tumor reduction or complete remission) is modified in this investigation to adapt to the more general clinical situation. We define good responders as patients under the new treatment, who benefit from it. Their benefit is manifested in the fact that their value of the survival time is longer than that of patients with the same characteristics (predictive factors), randomized in the control group. The causal inference approach implies a comparison between what actually happened with the new treatment and what would have happened if the patient had received the control treatment. The key challenge is that it is not possible to observe both outputs simultaneously in the same patient and should be approximated with reference to a comparison group. In the methodology proposed by Alonso (2015), the authors introduce a sensitivity analysis to handle this problem. They assume a range of possible values for the correlation between the potential outcomes and for each correlation they estimate the probability of treatment success for an individual patient. Considering this, here an individual is classified as a good responder if all their probabilities of treatment success are estimated greater than 0.5. We define bad responders to be patients under the new treatment who are harmed by it, that is, if all the probabilities of treatment success are lower than 0.5. Consequently, rare-responders would be patients who are neither good nor bad responders. In this last group are the patients that depending on the correlation between the potential outcomes can have values of probability of treatment success above and below 0.5

*Subgroup analyses for survival benefit*

To show the heterogeneity in the response to CIMAvax-EGF, the Kaplan Meier survival estimation was carried out in the good and poor responder patients. The log-rank test was used to compare the survival for the treated and control groups inside the subgroups identified by the biomarkers.

## III. RESULTS

*Calculation of the predictive causal inference association for all possible models*

Predictive individual causal association was univariate evaluated for each predictor, along all possible "realities" for the correlation between potential outcomes. The mean, minimum and maximum values of PCI for each model, as well as the accuracy is shown in Table 1. Note that all univariate models were negligible or low accurate. The proportion of CD4+ cell was the best predictor in the univariate analysis with a PCI of 0.49.

**Table 1**. Predictive Individual Causal Inference association (PCI) for each predictor.

| Predictors | PCI mean (min-max) |
|---|---|
| Basal EGF concentration | 0.005 (0.003-0.008) |
| Eosinophils | 0.001 (0.001-0.002) |
| Lymphocytes | 0.007 (0.005-0.012) |
| Neutrophils | 0.030 (0.019-0.051) |
| Platelets | 0.036 (0.023-0.059) |
| Monocytes | 0.163 (0.108-0.259) |
| NLR | 0.025 (0.016-0.043) |
| PLR | 0.004 (0.003-0.008) |
| Proportion of CD19+ B cell | 0.053 (0.034-0.087) |
| Proportion of CD8+ T cell | 0.087 (0.062-0.127) |
| Proportion of CD8-CD28- T cell | 0.148 (0.098-0.239) |
| CD4+/CD8+ ratio | 0.443 (0.324-0.626) |
| Proportion of CD4+ T cell | 0.486 (0.353-0.694) |

*Selection of a model taking into account its complexity and prediction accuracy*

The PCI values were calculated for the 8204 models defined considering all possible combinations of the biomarkers. The model with the 13, biomarkers reach a mean PCI value of 0.98 indicating that prediction accuracy of the complete model is very high. Figure 1 shows the increase in the values of PCI with the number of predictors. Note that from the model with 5 predictors (Proportion of CD4+ T cell, basal EGF concentration, NLR, Monocytes and Neutrophils) a high accuracy is obtained. Importantly, the minimum PCI obtained with the 5-dimensional predictor (min= 0.74) already exceeds the maximum value obtained for the best univariate predictor (Proportion of CD4+ T cell, max Pred-ICA=0.69). This was the model selected.

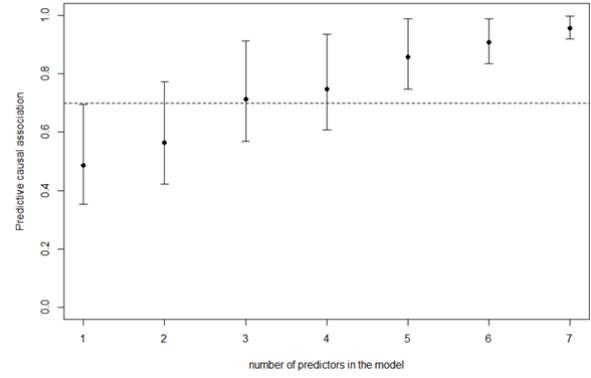

**Figure 1.** Predictive individual causal association by the best model according to the number of predictors: 1-proportion of CD4+ T cell, 2- proportion of CD4+ T cell and absolute monocytes counts, 3- proportion of CD4+ T cell, NLR and Neutrophils, 4- proportion of CD4+ T cell, NLR, Neutrophils and Eosinophils, 5- Proportion of CD4+ T cell, basal EGF concentration, NLR, Monocytes and Neutrophils, 6- Proportion of CD4+ T cell, Proportion of CD8+ T cell, basal EGF concentration, NLR, Monocytes and Neutrophils, 7- Proportion of CD4+ T cell, Proportion of CD8+ T cell, basal EGF concentration, NLR, Monocytes, Neutrophils and Eosinophils

*Identification of good, rare and bad responders to the treatment*

Using the selected model, we can compute the probability of treatment success for an individual patient. These probabilities are shown in Figure 2 in three different scenarios of predictor's values. The probability of treatment success is higher than 0.5 in all "realities" for the first imaginary patient, an individual with basal EGF concentration=1700, CD4+ T cells=65, CD4/CD8 ratio=3, NLR=2 and Neutrophils=55. This probability increases when the value of the correlation between the potential effect of the treatment and the effect of best supportive care (control) increases. This patient may be considered as a good responder to the treatment. For the second scenario, we considered a patient with basal EGF concentration =800, CD4+ T cells =60, CD4/CD8 ratio=3, NLR=1, Neutrophils=70. For this patient the treatment has the same probability of success as of failure. Finally, for the last scenario, a patient with lower values of basal EGF concentration in serum, low proportion of CD4+ T cell, and low NLR (basal EGF concentration =500, CD4+ T cells =20, CD4/CD8 ratio=1, NLR=0.5, Neutrophils=75) was considered. The probability of treatment success in this case is lower than 0.5 in all possible "realities". This patient may be considered as a bad responder to the treatment. Using the Excel score sheet developed (see the supplement) one can calculate the expected individual causal treatment effect. Therefore, we classify all patients in three groups: good responders, rare and bad responders.

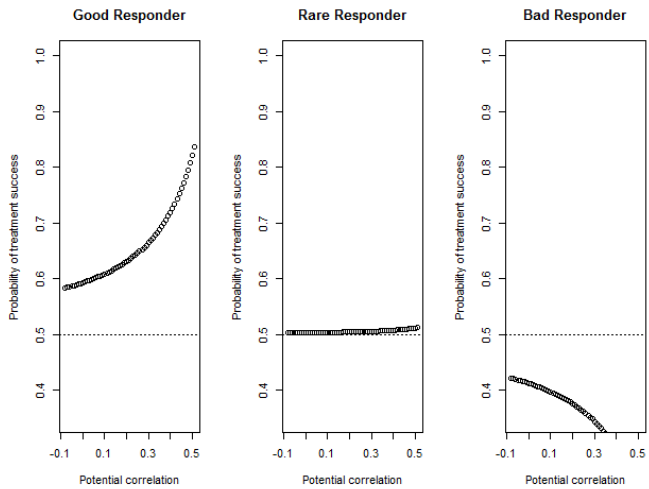

**Figure 2** Predictive probability of treatment success for three examples of a) Good responder (basal EGF concentration=1700, CD4+ T cells=65, CD4/CD8 ratio=3, NLR=2, Neutrophils=50), b) Rare (basal EGF concentration =900, CD4+ T cells =35, CD4/CD8 ratio=3, NLR=2, Neutrophils=55) and c) Bad responders (basal EGF concentration =200, CD4+ T cells =10, CD4/CD8 ratio=1, NLR=1, Neutrophils=60) to CIMAvax-EGF

*Subgroup analyses for survival benefit*

The survival curves for good and bad responders is shown in the Figure 3. A great difference is observed between the treated and control groups but for patients classified as good responders according to the model. Almost 50% of these patients resulted long term survivors (live more than 2 years), while no long term survivors were observed between patient received best supportive cares only. In contrast, patients predicted as bad responders had a survival time comparable to the controls with the similar biomarkers characteristics.

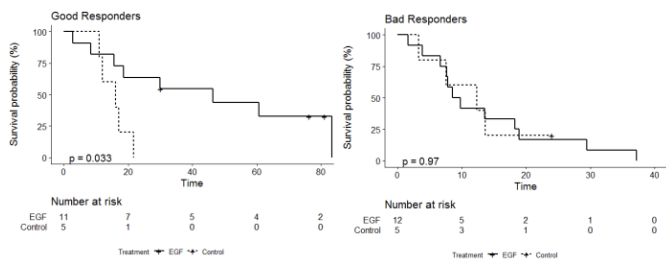

**Figure 3**. Kaplan Meier survival curves for patient treated with CIMAvax-EGF and control for a) good responders, b) poor responders.

## IV. DISCUSSION

Use e Together with the predictive biomarkers previously reported (T cell subpopulations and the EGF serum levels) our finding suggest, for the first time, the importance of the blood markers (NLR, Monocytes and Neutrophils) to predict the therapeutic success of CIMAvax-EGF. From the methodological point of view, the study shows the usefulness of the multivariate causal inference approach to identify a good combination of predictive biomarkers and to illustrate the application of this methodology for the identification of subgroups of advanced lung cancer patients with good, rare and bad probabilities of success with CIMAvax-EGF.

Previous studies evaluating the biomarkers of CIMAvaxEGF, used a univariate approach and looked at a single biological phenomenon. On the one hand, there are studies reporting the role of the EGF circulating in blood in the success of CIMAvaxEGF and the relationship with the mechanism of action of this immunotherapy [1, 3]. They highlighted that the EGF level in patients' sera could be simultaneously a biomarker of poor prognosis and a predictive factor of CIMAvax-EGF benefit. On the other hand, the biomarkers related to immunosenescence and its relationship with the CIMAvax-EGF therapeutic success was assessed by Saavedra and colleagues (2016). They found that patients treated with CIMAvax-EGF with CD4+ T cells counts greater than 40%, CD8+CD28− T cells counts lower than 24% and a CD4/ CD8 ratio >2 after first-line platinum-based chemotherapy, achieved a significantly large median survival, as compared to controls with the same phenotype. In these studies, the biomarkers, as is common in medical research, where dichotomized using the median or an optimal cut point. This follows the clinical practice of labelling individuals as having or not an attribute. Nevertheless, it is well known in the methodological literature that dichotomization of continuous variables introduces major problems include loss of information, reduction in power and uncertainty in defining the cut point [9]. The present study has as a strength that allows the analysis of the role of all these biomarkers jointly taking advantage of the continuous measurement scale. Moreover, it incorporates some markers of peripheral blood that have been related to the inflammatory process [10].

Currently, most pre-treatment predictors of therapeutic success are evaluated using correlational techniques. Regression model, the most used method, is able to include prognostic variables as a main effect and predictive variables in an interaction with treatment variable. A statistically significant and large interaction effect usually indicates potential subgroups that may have different responses to the treatment. However, in the conventional regression method to specify the interaction term the knowledge of predictive variables is required in advance. Such pre-specification of a regression model usually fails to identify the correct subgroups due to large number of covariates and complex interactions among them. The methodology used here was introduced to overcome these problems [8].

We recognize that there are limitations in the study because the small sample size and the possible biases inherent in any retrospective study. A new confirmatory study with a larger sample size is now being carried out to validate the predictive value of the biomarkers identified.